\documentstyle[12pt]{article}
\newcommand{\IP}{\relax{\rm I\kern-.18em P}}
\newcommand{\be}{\begin{equation}}
\newcommand{\ee}{\end{equation}}
\newcommand{\non}{\nonumber}
\newcommand{\bea}{\begin{eqnarray}}
\newcommand{\eea}{\end{eqnarray}}
\newcommand{\ba}{\begin{array}}
\newcommand{\ea}{\end{array}}

\newcommand{\pa}{\partial}
\newcommand{\hpa}{\hat{\partial}}

\newcommand{\de}{\delta}

\newcommand{\rar}{\rightarrow}

\newcommand{\hX}{\hat{X}}
\newcommand{\hF}{\hat{F}}

\newcommand{\hY}{\hat{Y}}

\newcounter{mycount}

\begin{document}
\pagestyle{empty}
\vspace{-1mm}
\begin{flushright}
NSF-ITP-97-026 \\
CERN-TH/97-54\\ G\"{o}teborg ITP 97-02 \\hep-th/9703195
\end{flushright}
\vspace{5mm}
\begin{center}
{\bf \LARGE Special Geometry and Automorphic Forms\\}
\vspace{8mm}
{\bf \large Per Berglund\footnote{berglund@itp.ucsb.edu}\\}
\vspace{2mm}
{Institute for Theoretical Physics \\ University of California \\
Santa Barbara, CA 93106, USA\\}
\vspace{4mm}
{\bf \large M{\aa}ns Henningson\footnote{henning@nxth04.cern.ch}\\}
\vspace{2mm}
{Theory Division, CERN\\ CH-1211 Geneva 23, Switzerland\\}
\vspace{4mm}
{\bf \large Niclas Wyllard\footnote{wyllard@fy.chalmers.se}\\}
\vspace{2mm}
{Institute of Theoretical Physics\\ S-412 96  G\"oteborg, Sweden\\}
\end{center}
\vspace{6mm}
\begin{abstract}
We consider the special geometry of the vector multiplet
moduli space in compactifications of the heterotic string on $K3 \times
T^2$ or the type IIA string on $K3$-fibered Calabi-Yau threefolds.
In particular, we
construct a modified dilaton that is invariant under $SO(2, n; {\bf
Z})$ $T$-duality transformations at the non-perturbative level and
regular everywhere on
the moduli space. The invariant dilaton, together with a set of other
coordinates that transform covariantly under $SO(2, n; {\bf Z})$,
parameterize the moduli space. The construction involves a meromorphic
automorphic function of $SO(2, n; {\bf Z})$, that also depends on the
invariant dilaton. In the weak coupling limit, the divisor of this
automorphic form is an integer linear combination of the rational
quadratic divisors where the gauge symmetry is enhanced classically. We
also show how the non-perturbative prepotential can be expressed in
terms of meromorphic automorphic forms, by expanding a $T$-duality
invariant quantity both in terms of the standard special coordinates
and in terms of the invariant dilaton and the covariant coordinates.
\end{abstract}
\vspace{5mm}
{March, 1997. \\ CERN-TH/97-54}
\newpage

\pagestyle{plain}
\setcounter{page}{1}
\setcounter{equation}{0}
\section{Introduction}
Theories with $N = 2$ extended supersymmetry in $d = 4$ space-time
dimensions have proven to be a remarkably fertile area of study,
exhibiting a rich variety of physical phenomena, while at the same
time being sufficiently constrained to be amenable to a quantitative
analysis. This has been used fruitfully in quantum field theory with
rigid supersymmetry as well as in supergravity
and
also in $N = 2$ string theory compactifications to $d = 4$.

In this paper, we will consider the theories that arise when,
for example, the heterotic string is compactified to four dimensions
on $K3 \times T^2$.
At the perturbative level, such a model is invariant under
$T$-duality transformations that form a group isomorphic to $SO(2, n;
{\bf Z})$ for some $n$ \cite{Giveon-Porrati-Rabinovici}. This symmetry
is seemingly explicitly broken
by the non-perturbative corrections, though. The classically invariant
dilaton
will transform in a non-trivial way under $T$-duality in the exact
theory. Furthermore, the exact holomorphic
prepotential $F$, that describes the
vector multiplet moduli space, is a transcendental
function (in so called special coordinates) with polylogarithm
singularities and complicated transformation properties under $SO(2,
n; {\bf Z})$.

It has been known for some time that at the perturbative level in the
heterotic string, one can introduce an invariant dilaton
\cite{deWit95},
which however is not a special coordinate. This construction involves a
certain automorphic form of $SO(2, n; {\bf Z})$. Furthermore, by taking
a suitable number of derivatives of the perturbative
contribution to $F$ one obtains a meromorphic automorphic
form of $SO(2, n; {\bf Z})$
\cite{deWit95,Antoniadis95,Antoniadis96,KLT,Harvey-Moore1}.
In this paper, we will generalize these ideas to the non-perturbative
level by constructing a distinguished
set of coordinates, that are not of the special coordinates type, but
transform in a simple way under $T$-duality. More precisely, we
construct a $T$-duality invariant dilaton and a set of coordinates that
transform covariantly under $SO(2, n; {\bf Z})$. In a sense, by going
to this
frame we compensate for the quantum corrections. To get an invariant
dilaton
which is regular everywhere on the moduli space, and thus can be used
as an expansion parameter, we introduce a meromorphic automorphic
function of $SO(2, n; {\bf Z})$, that also depends on the invariant
dilaton.
In the limit where the invariant dilaton is large, i.e. the weak
coupling limit, the divisor of this automorphic function is an integer
linear
combination of so called rational quadratic divisors \cite{Borcherds}.
These are the divisors where the heterotic string classically acquires
an
enhanced non-abelian gauge symmetry and the perturbative description of
the
type II string breaks down. The invariant dilaton and the covariant
coordinates may be used to express the non-perturbative prepotential in
terms of meromorphic automorphic forms of $SO(2, n; {\bf Z})$. The idea
is to expand some $T$-duality invariant quantity both in terms of the
standard special coordinates and in terms of the invariant dilaton and
the covariant coordinates. Such formulas were given for a particular
two-parameter model in \cite{Henningson1}, but without the geometric
interpretation discussed in this paper.

We see several possible applications of the geometric structure
discussed in this paper. It is clearly related to the problem of
understanding
the heterotic string at a non-perturbative level. Using the type IIA
picture, we get a relationship between the number of rational curves
on $K3$-fibered Calabi-Yau threefolds and automorphic forms that
would be interesting to study further. On the type IIB side, the
Picard-Fuchs equations for the periods of the mirror manifold of a
$K3$-fibered Calabi-Yau threefold \cite{Klemm95,Lian-Yau}
should simplify in the covariant coordinates. This is probably related
to the problem of finding a criterion characterizing the mirror
manifolds of
$K3$-fibered Calabi-Yau threefolds. More speculatively, it would be
nice if
these ideas generalize to $K3$-fibered Calabi-Yau spaces of higher
dimension and their mirrors.

The meromorphic automorphic function that arises in the construction of
the
perturbatively invariant dilaton has remarkable mathematical
properties. Its divisor carries information about the spectrum
of massless particles. Furthermore, it can be written as an infinite
product,
where the exponents are the Fourier coefficients of an $PSL(2; {\bf
Z})$
modular form \cite{Harvey-Moore1,Borcherds,Henningson2}.
(The latter form is essentially the elliptic genus of the $K3$
non-linear sigma model with an $E_8 \times E_8$ or $S\hspace{-0.5mm}pin
(32) /
{\bf Z}_2$ vector bundle that defines the heterotic compactification).
Most
importantly, this form is the denominator formula for a generalized
Kac-Moody algebra that has been conjectured to be a broken gauge
symmetry algebra in string theory \cite{Harvey-Moore1}.
The gauge bosons of this algebra are the BPS saturated states,
which naturally form an algebra \cite{Harvey-Moore2}. It
would be very interesting to give a non-perturbative generalization
of these statements. In the type II picture, it would be desirable
to relate the properties of the automorphic forms that arise in this
construction to the data specifying the $K3$-fibered Calabi-Yau
threefold for example as a subvariety of a toric variety.

This paper is organized as follows: In the next section, we give a
quick
review of theories with $N=2$ supersymmetry in $d=4$ space-time
dimensions,
with particular emphasis on special geometry of the type that
arises in heterotic compactifications on $K3\times T^2$ and type IIA
compactifications
on $K3$-fibered Calabi-Yau threefolds. In section three we discuss
these two
types of compactifications in more detail. In section four, we discuss
the singularity structure of the vector multiplet moduli space. In
section five, which is the main part of the paper, we introduce the
covariant coordinates. These are used in section six to show how the
prepotential can be expressed in terms of automorphic forms. Finally,
in the appendix, we discuss some cases with a small number of moduli
more explicitly.

\setcounter{equation}{0}
\section{Review of $N=2$ in $d=4$}

At low energies, theories with $N = 2$ supersymmetry in $d = 4$
space-time dimensions are described by some effective
Lagrangian governing the dynamics of the massless multiplets.
The massless spectrum consists of:
\begin{enumerate}
\item
Exactly one supergravity multiplet (in the case of theories with
local supersymmetry), consisting of a spin 2 graviton, two spin 3/2
gravitini and a spin 1 graviphoton.
\item
$N_V$ vector multiplets, each consisting of a spin 1
gauge field, two spin 1/2 gaugini, and two spin 0 fields. These
fields transform in the adjoint of the gauge group $G$ of the theory.
\item
$N_H$ hyper multiplets, each consisting of two spin
1/2 quarks and four spin 0 fields. These fields transform in some
representation $R \oplus
\bar{R}$ of $G$
\end{enumerate}
A most important point is that the superpotential for the scalar
fields in the vector and hyper multiplets has flat directions, so we
have a moduli space ${\cal M}$ of inequivalent vacua. Furthermore,
the couplings between vector and hyper multiplets of this Lagrangian
are constrained to be a supersymmetric extension of `minimal
coupling' \cite{deWit85}. In particular, there are no couplings between
neutral hyper multiplets and vector multiplets. It follows that, at
least
locally, the moduli space factorizes as
\be
{\cal M} \cong {\cal M}_V \times {\cal M}_H,
\ee
where the two factors ${\cal M}_V$ and ${\cal M}_H$ are parameterized
by the vacuum expectation values of the scalars in the vector and
hyper multiplets respectively. The hyper multiplet moduli space
${\cal M}_H$ has the structure of a quaternionic manifold
\cite{Bagger-Witten}. In this paper, we will focus on the vector
multiplet
moduli space ${\cal M}_V$, which is subject to the constraints of
special
geometry \cite{deWit84,deWit85,Cremmer-etal}. The implications of these
constraints will be reviewed below.

A universal modulus in string theory is the dilaton-axion $\theta /
2 \pi + i 4 \pi / g^2$, where $g$ and $\theta$ are the coupling
constant and the theta angle respectively. This modulus organizes the
perturbative and non-perturbative quantum corrections. It is therefore
important to know to which type of multiplet it belongs. We can get $N
= 2$ local supersymmetry in $d = 4$ space-time dimensions by
compactifying
\begin{enumerate}
\item
A string theory with $N = 2$ supersymmetry in $d = 10$ dimensions,
i.e. the type IIA or type IIB string, on a manifold of $SU(3)$
holonomy, i.e. an arbitrary Calabi-Yau threefold. The dilaton-axion
is then part of a hyper multiplet \cite{Seiberg}.
\item
A string theory with $N = 1$ supersymmetry in $d = 10$ dimensions,
i.e. the $E_8 \times E_8$ or $S\hspace{-0.5mm}pin(32)/{\bf Z}_2$
heterotic
string or
the type I string, on a manifold of $SU(2)$ holonomy, i.e. $K3 \times
T^{2}$. In the heterotic case, the dilaton-axion is part of a vector
multiplet\footnote{More precisely, the dilaton-axion belongs
to a vector-tensor
multiplet, which on-shell is dual to a vector multiplet
\cite{deWit95}.},
whereas it is a combination of vector and
hyper multiplet moduli in the type I case \cite{Vafa95}.
\end{enumerate}
There are also more general possibilities such as, for instance,
M-theory
compactified on $CY_{3}\times S^{1}$, or F-theory compactified on
$CY_{3}\times T^2$. The former vacua are dual to type IIA theory on the
same  Calabi-Yau threefold while
the latter is dual to the heterotic string on $K3 \times T^2$.
Note however,
that while in the case of M-theory there is no condition on the
manifold, F-theory requires that the Calabi-Yau is an elliptic
fibration.
In this paper, we will restrict our study to the special geometry that
arises in
compactifications of the heterotic string on $K3 \times T^{2}$ or the
type IIA string on $K3$-fibered Calabi-Yau threefolds. In many cases,
these models are dual to each other \cite{Kachru-Vafa,FHSV,Aspinwall}.
One can also use mirror symmetry
to relate them to compactifications of the type IIB string on mirror
manifolds of $K3$-fibered Calabi-Yau threefolds
\cite{mirror_manifolds}.
The geometry of ${\cal M}_V$ is encoded in a holomorphic function $F$
called the prepotential. In the type II case, $F$ can be calculated
exactly at string tree level by going to a weak coupling limit.
For type IIA,  ${\cal M}_V$ can be identified with
the complexified K\"ahler cone of the Calabi-Yau threefold $X$. In
the large radius limit, $F$ is essentially the triple intersection
form on $H^{1, 1} (X; {\bf Z})$ \cite{Strominger-Witten}, but
world-sheet
instantons, i.e. rational curves on $X$, give corrections at finite
radius \cite{DSWW}. In principle these instanton corrections can be
computed directly on the type IIA side \cite{AspMor}. However, the
calculation
is more tractable by going to the mirror manifold $\tilde{X}$ and the
corresponding type IIB theory.
For
type IIB, ${\cal M}_V$ can be identified with the complex structure
moduli space of the Calabi-Yau threefold on which the theory is
compactified. $F$ can then be
determined exactly at world-sheet tree level by going to a large
radius limit. The computation amounts to determining the periods
of the holomorphic, nowhere vanishing section of the canonical bundle
of $\tilde{X}$.  The prepotential is expressed in terms of these
periods.
By using the mirror map, relating the periods to the so called
special coordinates, we obtain $F$ for the type IIA theory.
In the
heterotic case, $F$ is classically given by the metric on the Narain
moduli space of the two-torus, but it receives perturbative (one-loop
only) and non-perturbative quantum corrections controlled by the
dilaton-axion. The non-perturbative corrections can be attributed to
space-time instantons.

\subsection{Review of Special Geometry}
The constraints of special geometry amount to the following
\cite{Strominger90}: ${\cal M}_V$ is a K\"ahler manifold of restricted
type\footnote{Also called a Hodge manifold.}, i.e. the cohomology class
of the K\"ahler form is an even element of
$H^2({\cal M}_V; {\bf
Z})$. This determines a line bundle ${\cal L}$ over ${\cal M}_V$ the
first Chern class of which equals the K\"ahler class of ${\cal M}_V$.
Furthermore, there exists a flat $Sp(4 + 2 n; {\bf R})$ holomorphic
vector bundle ${\cal H}$ over ${\cal M}_V$ with a compatible
hermitian metric where $n = {\rm dim}_{\bf C} {\cal M}_V - 1$, and a
holomorphic section $\Pi$ of ${\cal L} \otimes {\cal H}$ such that
the K\"ahler potential $K$ of ${\cal M}_V$ can be written as
\be
K = - \log i \left( \bar{\Pi}^t J \Pi \right) ,
\ee
where $J$ is the $Sp(4 + 2 n)$ invariant metric. In a more general
setting
further conditions are needed to define special geometry
\cite{Craps97}.
We work in a particular set of homogeneous special
coordinates\footnote{Indices
$I, J, \ldots$ always take the values $0, 1,
\ldots, n + 1$.} $X^I$, which are functions defined on ${\cal M}_V$,
for which
the section $\Pi$ can be written in the form
\be
\Pi^t = \left( X^0, X^1, \ldots, X^{n + 1} ; F_0, F_1, \ldots, F_{n +
1} \right) ,
\ee
where $F_I = \pa_I F = {\pa F \over \pa X^I}$ and the prepotential
$F$ is a function of the $X^I$ of degree two in $X^I$. In this basis,
the $Sp(4 + 2 n)$ invariant metric is
\be
J = \pmatrix{ 0 & I \cr - I & 0 \cr} ,
\ee
where $I$ is the $(2 + n) \times (2 + n)$ unit matrix. The
homogeneous special coordinates arise naturally in supergravity: The
$X^\alpha$ for $\alpha = 1, \ldots, n + 1$ are the complex scalars of
the vector multiplets, and $X^0$ is a non-dynamical complex scalar
corresponding to the graviphoton. As coordinates on ${\cal M}_V$ we
can then take the inhomogeneous special coordinates $X^\alpha / X^0$
for $\alpha = 1, \ldots, n + 1$.

In general, ${\cal M}_V$ has singularities along certain divisors
where additional fields become massless. If we encircle such a
singularity, the section $\Pi$ will undergo a monodromy
transformation of the form $\Pi \rightarrow M \Pi$, where $M \in Sp(4
+ 2 n; {\bf Z})$, i.e. $M^t J M = J$. These transformations
generate the monodromy group of the theory. Note that the K\"ahler
potential $K$ is invariant under such a transformation.

\setcounter{equation}{0}
\section{The cases with $S$- and $T$-duality}
We will be interested in the particular cases corresponding to the
compactifications of
\begin{enumerate}
\item
The $E_8 \times E_8$ or $S\hspace{-0.5mm}pin(32)/{\bf Z}_2$ heterotic
string on
$K3
\times T^{2}$.
\item
The type IIA string on a $K3$-fibered Calabi-Yau threefold.
\item
The type IIB string on the mirror manifold of a $K3$-fibered
Calabi-Yau threefold.
\end{enumerate}
These compactifications are conjectured to be related by
heterotic/type IIA duality and mirror symmetry between type IIA and
type IIB.

The special geometry of the corresponding vector multiplet moduli
space ${\cal M}_V$ can be characterized as follows: We work with
a particular set of homogeneous special coordinates $X^I$. The
corresponding inhomogeneous special coordinates, that parameterize
${\cal M}_V$, are denoted $S = X^1 / X^0$ and\footnote{Indices $i, j,
\ldots$ always take the values $2, \ldots, n + 1$.}
$T^i = X^i / X^0$. Following standard practice, it is
convenient to introduce the period vector $\hat{\Pi}$, which is
related to $\Pi$ by an $Sp(4 + 2 n; {\bf Z})$ transformation, as
\bea
\hat{\Pi}^t & = & \left( \hX^0, \hX^1, \hX^2, \ldots, \hX^{n + 1} ;
\hF_0, \hF_1, \hF_2, \ldots, \hF_{n + 1} \right) \cr
& = & \left( X^0, F_1, X^2, \ldots, X^{n + 1} ; F_0, - X^1, F_2,
\ldots, F_{n + 1} \right) .
\eea
We also define the symmetric tensor $\eta_{IJ}$ by $\eta_{01} = - {1
\over 2}$, $\eta_{00} = \eta_{11} = \eta_{0i} = \eta_{1i} = 0$ and
$\eta_{ij} = {\rm diag} ( -1, 1, \ldots, 1)$. The extra
condition that we impose on the theory is that the monodromy group
contains a subgroup generated by the elements of a group isomorphic to
${\bf Z} \times SO(2, n; {\bf Z})$,
where we refer to the two factors as the $S$- and $T$-duality groups
respectively. The $S$-duality group acts as
\bea
\hX^I & \rightarrow & \hX^I \cr
\hF_I & \rightarrow & + \hF_I + 2 b \eta_{IJ} \hX^J, 
\label{theta_shift}
\eea
where $b \in {\bf Z}$. An element $U$ of the $T$-duality group, defined
as the group of matrices $U^I{}_J$ such that $(U^t)_I{}^K \eta_{KL}
U^L{}_J = \eta_{IJ}$, acts as
\bea
\hX^I & \rightarrow & U^I{}_J \hX^J \cr
\hF_I & \rightarrow & (U^{t - 1})_I{}^J \left(\hF_J +
(\Lambda_U)_{JK} \hX^K \right) , \label{T-duality}
\eea
where $(\Lambda_U)_{JK}$ is some symmetric, integer-valued matrix. We
will discuss the form of the $\Lambda_U$-matrices later. Finally, we
require $F_1$ to be regular as $X^1 / X^0 \rightarrow i \infty$ with
$X^0$ and the $X^i$ held fixed.

The general solution to the conditions on the prepotential following
from the $S$-duality transformation (\ref{theta_shift}) is
\be
F = {X^1 \over X^0} X^i \eta_{ij} X^j + f, \label{prepotential}
\ee
where the function $f$ is of degree two in $X^I$, invariant under
$X^1 \rightarrow X^0 + X^1$, and regular in the $X^1 / X^0
\rightarrow i \infty$ limit. This means that $f$ can be expanded in a
Fourier series as
\be
f = \sum_{k = 0}^\infty q^k f_k , \label{Fourier}
\ee
where $q = e^{2 \pi i S}$ and the functions $f_k$ are
independent of $X^1$ and of degree two in $X^0$ and the $X^i$. The
$T$-duality transformations impose some constraints on these
functions, which we will analyze in detail later.

\subsection{The heterotic picture}
For the case of a compactification of the $E_8 \times E_8$ or
$S\hspace{-0.5mm}pin(32)/{\bf Z}_2$ heterotic string on $K3 \times
T^{2}$ the
interpretation of the above structure is as follows: The modulus $S$
is the dilaton-axion $\theta / 2 \pi + i 4 \pi / g^2$, and the $T^i$
parameterize the moduli space of the torus and a flat $U(1)^{n - 2}$
bundle over it, i.e. $T = T^2-T^3$ and $U = T^2+T^3$ are the complex
structure and complexified K\"ahler moduli respectively and $V^a =
T^a$ for $a = 4, \ldots, n + 1$ are the Wilson line moduli
\cite{Narain}.
The moduli of the $K3$ surface and the $E_8 \times E_8$ or
$S\hspace{-0.5mm}pin(32)/{\bf
Z}_2$ bundle parameterize the hyper multiplet moduli space ${\cal
M}_H$. The dimension of this space may jump at particular points in
${\cal M}_V$ where additional hyper multiplets become massless. By
`Higgsing', i.e. giving vacuum expectation values to such hyper
multiplets, we may give masses to some vector multiplets and thus
change the number $n$ in the interval $0 \leq n \leq 18$.
Classically the vector multiplet moduli space ${\cal M}_V$ is a
direct product:
\bea
{\cal M}^{\rm classical}_V & \cong & {\cal M}^{\rm dilaton} \times
{\cal M}^{\rm Narain} \cr
& \cong & ({\bf Z} \backslash {SU(1, 1) \over U(1)}) \times
(SO(\Gamma^{2, n}) \backslash {SO(2, n) \over SO(2) \times SO(n)} ) ,
\label{product_space}
\eea
where $\Gamma^{2, n}$ is the Narain lattice of the indicated signature.
Spaces that locally are
of the form (\ref{product_space}) are in fact the only special
geometries that can be written as direct products \cite{Ferrara89}.
The prepotential of such a space is precisely given by the first term
in
(\ref{prepotential}). Quantum mechanically, the geometry receives
perturbative and non-perturbative corrections that are encoded in the
second term in (\ref{prepotential}). In the expansion
(\ref{Fourier}), the $q^k f_k$ term is the one-loop contribution in
the $k$-instanton sector. Note that $q$ is precisely the exponential
of the instanton action. In particular, the $f_0$ term is thus the
perturbative contribution.
At the semi-classical level, where $f_k = 0$ for $k > 0$, the theory
is in fact invariant under continuous real shifts of $S$ and this
symmetry is explicitly broken down to the discrete Peccei-Quinn
subgroup
(\ref{theta_shift}) by instantons just as expected. We expect the
$SO(\Gamma^{2, n})$ $T$-duality group to survive in the exact theory.
Indeed, it can be interpreted as a discrete gauge symmetry and
should therefore not be explicitly broken by any anomalies \cite{DHS}.

\subsection{The type II picture}
Next we turn to the case of compactification of the type IIA string
on a Calabi-Yau threefold $X$. The hyper multiplet moduli space
${\cal M}_H$ is then $4(h^{2, 1}(X) + 1)$ real-dimensional and is
parameterized by the complex structure of $X$, the moduli arising from
compactifications of the Ramond-Ramond three-form, and the
dilaton-axion.
The vector multiplet moduli space ${\cal M}_V$ is $h^{1, 1}(X)$
complex-dimensional and can be identified with the complexified
K\"ahler cone of $X$. Using the isomorphism $H^2(X; {\bf Z}) \cong
H_4(X; {\bf Z})$ we can associate divisors $D_1$ and $D_i$ in the
latter group to the moduli $T^1 = S$ and the $T^i$. The prepotential
$F$ can then be written in the form
\be
F = - {1 \over 3!} (D_\alpha \cdot D_\beta \cdot D_\gamma) T^\alpha
T^\beta T^\gamma + {i \zeta (3) \over 16 \pi^3} \chi (X) + {1 \over
(2 \pi i)^3} \sum_R n(R) {\rm Li}_3 \left( \exp 2 \pi i R_\alpha
T^\alpha \right) , \label{curvcount}
\ee
where $D_\alpha \cdot D_\beta \cdot D_\gamma$ are the triple
intersection numbers of $X$, $\chi (X)$ is the Euler characteristic
of $X$, and $n(R)$ is the number of rational curves of multi degree
$R = (R_1, \ldots R_{n + 1})$ on $X$. The mirror hypothesis states the
equivalence of the
compactifications of the type IIA string on $X$ and the type IIB
string on the mirror manifold $\tilde{X}$ of $X$. The data entering in
(\ref{curvcount}) can in many cases be effectively computed by
considering the complex structure moduli space of $\tilde{X}$
\cite{mirror_manifolds}. Comparing (\ref{curvcount}) with
(\ref{prepotential}) and (\ref{Fourier}), we see that the divisor
$D_1$ squares to zero, and that $C \cdot D_1 \geq 0$ for all curves
$C$ in $X$. The existence of such a nef divisor of numerical
$D$-dimension equal to one implies that $X$ is a fiber bundle over
$\IP^1$ with the generic fiber being a $K3$-surface \cite{Oguiso}.
Thus a necessary condition for the compactification of the type IIA
theory on $X$ to be dual to a compactification of the heterotic
string on $K3 \times T^{2}$ is that $X$ is $K3$-fibered
\cite{Klemm95,Aspinwall}.
A point on the $\IP^1$ base times the
generic $K3$ fiber equals the divisor $D_1$ when considered as an
element of $H_4 (X;{\bf Z})$. If we neglect contributions to
$H_4 (X; {\bf Z})$ from
reducible `bad' fibers, the remaining generators are given by the
fundamental class of the $\IP^1$ base times algebraic two-cycles
in the generic $K3$ fiber that are invariant under the monodromy
around closed curves on the base. The lattice of such two-cycles
equals the monodromy invariant part of the Picard lattice ${\rm
Pic}(D_1)$ of the generic $K3$ fiber, and has rank $n$ in the
interval $1 \leq n \leq \rho (D_1)$. Here  $\rho (D_1)$ is the Picard 
number of the generic $K3$ fiber, and $1 \leq \rho (D_1) \leq 20$.
The modulus $S$ is the complexified K\"ahler form on the $\IP^1$ base. 
Shifting the $B$-field of the complexified K\"ahler form on the $\IP^1$
base by an element of $H^2(\IP^1; {\bf Z}) \cong {\bf
Z}$ is physically trivial, so $S$-duality should be valid in the exact
theory. Furthermore, the moduli $T^i$ parameterize the moduli
space of the complexified K\"ahler form on the generic $K3$ fiber.
This space is isomorphic to the second factor in
(\ref{product_space}),
where $\Gamma^{2, n}$ now is the monodromy invariant part of the
`quantum Picard lattice', i.e. the direct sum of ${\rm Pic}(D_1)$ and
the
hyperbolic plane $U = \pmatrix{0 & 1 \cr 1 & 0}$. Again we expect the
$T$-duality group to be a subgroup of the monodromy group in the exact
theory. We have thus found completely analogous structures
for the two different string theory realizations of these special
geometries.

\setcounter{equation}{0}
\section{The singularity structure}
We now turn to the singularities of the vector multiplet moduli space
${\cal M}_V$. The subspace of ${\cal M}_V$ at $q=0$, i.e. in the
$S \rightarrow i \infty$ limit,
takes the form of the second factor in (\ref{product_space}),
where $\Gamma^{2, n}$ is either the Narain lattice or the monodromy
invariant part of the quantum Picard lattice. This space can be
regarded as the Grassmannian of space-like two-planes in $\Gamma^{2,
n} \otimes {\bf R}$ modulo automorphisms of $\Gamma^{2, n}$. Let
$\alpha \in \Gamma^{2, n}$ be a root, i.e. $\alpha \cdot \alpha =
-2$. We define the rational quadratic divisor associated to $\alpha$
to be the subspace of the Grassmannian where $\alpha$ is in the
orthogonal complement of the space-like two-plane \cite{Borcherds}. 
In the heterotic
picture, this means that we have a purely left-moving vector with
length squared equal to two in the Narain lattice, i.e. we are at an
enhanced symmetry point where the unbroken gauge group acquires an
$SU(2)$ factor and some number $N_f$ of hyper multiplet doublets of
$SU(2)$
become massless. Larger non-abelian groups arise at the intersections
of rational quadratic divisors. In the type IIA picture, we use the
fact that a root $\alpha$ of the Picard lattice corresponds to a
rational curve, i.e. an embedded $\IP^1$. On the associated
rational quadratic divisor, $\alpha$ is orthogonal to the space-like
two-plane spanned by the complexified K\"ahler form of the generic
$K3$ fiber, i.e. the volume of and $B$-flux through this $\IP^1$
vanish. The perturbative analysis of the theory thus breaks down,
and we expect an $SU(2)$ factor in the gauge group together with a
number $N_f$ of massless hyper multiplet doublets
\cite{Witten,Aspinwall95}.

The massless spectrum on the rational quadratic divisors determines
the singularities of the functions $f_k$ in (\ref{Fourier}). This is
most easily understood by comparing the heterotic picture with the
corresponding $N = 2$ supersymmetric $SU(2)$ Yang-Mills theory with
$N_f$ massless hyper multiplet doublets. The one-loop beta function
is proportional to $4 - N_f$, so asymptotic freedom and non-trivial
infrared dynamics gives the restriction $N_f = 0, 1, 2, 3$. The
low-energy effective gauge couplings, given by the matrix of
second-derivatives of the prepotential, diverge as we approach the
divisor where $SU(2)$ is restored. In particular, the perturbative
contribution has a logarithmic divergence proportional to $4 - N_f$.
The contribution from the $k$-instanton sector has a pole of order
$(4 - N_f) k$ \cite{Seiberg-Witten}.

These results determine the singularities of the
functions $f_k$ in the string theory prepotential as we approach a
rational quadratic divisor $D(X) = 0$ where a $U(1)$ factor gets
enhanced to $SU(2)$ and $N_f$ hyper multiplet doublets become
massless \cite{Kachru1}. The function $f_0$ has a trilogarithm
singularity such that
\be
\pa_I \pa_J F = {1 \over 2 \pi i}
(4 - N_f) (\Omega_D)_{IJ} \log D(X) + {\rm regular}
\label{Omega}
\ee
for some integer-valued symmetric matrix $(\Omega_D)_{IJ}$, the precise
form of which depends on which $U(1)$ factor gets enhanced to
$SU(2)$. The $f_k$ for $k > 0$ behave like
\be
f_k \sim \left( D(X) \right)^{2 - k (4 - N_f)} .
\ee

The matrices $\Lambda_U$ for $U \in SO(2, n; {\bf Z})$ that appear in
the $T$-duality transformations (\ref{T-duality}) are also
constrained by the massless spectrum on the rational quadratic
divisors. To analyze these constraints, it is convenient to regard
the heterotic monodromy group as the fundamental group $\pi_1 ({\cal
M}_V)$ of the vector multiplet moduli space ${\cal M}_V$
\cite{Antoniadis95}. We have
already mentioned the fact that the classical vector multiplet moduli
space ${\cal M}_V^{\rm classical}$ is of the form
(\ref{product_space}). It follows that $\pi_1 \left({\cal M}_V^{\rm
classical} \right) \cong {\bf Z} \times SO(\Gamma^{2, n})$. To get
the true vector multiplet moduli space ${\cal M}_V$, we have to
remove the singularity locus given by the rational quadratic divisors
before modding out by $SO(\Gamma^{2, n})$, and this change affects
the fundamental group $\pi_1 ({\cal M}_V)$. This group can still be
regarded as being generated by the elements of $SO(\Gamma^{2, n})$,
but these generators no longer fulfill the relations of $SO(\Gamma^{2,
n})$. The reason is that a sequence of transformations that would be
equivalent to the identity element of $SO(\Gamma^{2, n})$ and thus of
$\pi_1 \left({\cal M}_V^{\rm classical} \right)$ might amount to
encircling a rational quadratic divisor $D(X) = 0$ and thus gives a
non-trivial element of $\pi_1 ({\cal M}_V)$. Indeed, under such a
transformation the prepotential $F$ would be shifted by ${1 \over 2}
(4 - N_f) \hX^I (\Omega_D)_{IJ} \hX^J$, where $(\Omega_D)_{IJ}$ is the
matrix in (\ref{Omega}). The induced transformation of the period
vector $\hat{\Pi}$ is
\bea
\hX^I & \rightarrow & \hX^I \cr
\hF_I & \rightarrow & \hF_I + (4 - N_f) (\Omega_D)_{IJ} \hX^J .
\eea
The requirements that the $T$-duality transformations
(\ref{T-duality}) form a group isomorphic to $\pi_1 ({\cal M}_V)$ and
that encircling a rational quadratic divisor induces the above
transformation gives a set of equations relating linear combinations
of the $\Lambda_U$-matrices to the $\Omega_D$ matrices. The general
solution to this system can be written in the form
\be
\Lambda_U = \Lambda_U^{\rm part} + \Lambda_U^{\rm hom} ,
\ee
where $\Lambda_U^{\rm part}$ is some particular solution and
$\Lambda_U^{\rm hom}$ is the general solution to the corresponding
homogeneous equations, i.e. the equations with all the
$\Omega$-matrices equal to zero. The latter equations read $(d^{(1)}
\Lambda^{\rm hom})_{U_1, U_2} = 0$ for all $U_1, U_2 \in SO(2, n;
{\bf Z})$, where
\be
(d^{(1)} \Lambda^{\rm hom})_{U_1, U_2} = \Lambda^{\rm hom}_{U_1 U_2}
- \Lambda^{\rm hom}_{U_2} - U_2^t \Lambda^{\rm hom}_{U_1} U_2 .
\ee
The interpretation is that in the absence of the singularities on
the rational quadratic divisors described by the $\Omega_D$-matrices,
the monodromy group is isomorphic to $SO(2, n; {\bf Z})$, and
performing first a $U_2$ transformation and then a $U_1$
transformation is then equivalent to performing a $U_1 U_2$
transformation. Adding a physically trivial term $\hX^I M_{IJ}
\hX^J$ to the prepotential,
where $M_{IJ}$ is an arbitrary symmetric matrix, would change
the $\Lambda_U$-matrices by $(d^{(0)} M)_U$, where
\be
(d^{(0)} M)_U = U^t M U - M .
\ee
Noting that $(d^{(1)} d^{(0)} M)_{U_1, U_2} = 0$ for all $U_1, U_2
\in SO(2, n; {\bf Z})$, we see that we are really interested in the
cohomology group
\be
H^1 = {\rm Ker}(d^{(1)}) / {\rm Im}(d^{(0)}) .
\ee
This group is trivial for $SO(2, 1; {\bf Z})$ and $SO(2, 2; {\bf
Z})$ \cite{Antoniadis95}, and we believe this to be true in the general
case of $SO(2, n; {\bf Z})$ as well.
The matrices $\Lambda_U$ for $U \in SO(2, n; {\bf Z})$ would then be
uniquely determined by the spectrum of massless particles on the
various rational quadratic divisors.

\setcounter{equation}{0}
\section{The invariant dilaton and the covariant coordinates}
The functions $f_k$ appearing in (\ref{Fourier}) are constrained by
the $T$-duality transformations (\ref{T-duality}). These constraints
can be worked out explicitly, but they are fairly complicated, due to
quantum corrections to the transformation laws of the
homogeneous special coordinates $X^I$
\cite{Henningson1}. Our aim in this section is to
find a new set of homogeneous coordinates $Y^I$, that are not special
coordinates, but which have the virtue that they transform in a simpler
way under $SO(2, n; {\bf Z})$.
If we then express some quantity which is invariant
under $SO(2, n; {\bf Z})$ in these new coordinates, we get a
set of functions related to the $f_k$ but with simpler transformation
properties.

Our strategy is now as follows: We will try to construct a set of
quantities $\hY^I$ that are invariant under the $S$-duality
transformations (\ref{theta_shift}), transform as
\be
\hY^I \rightarrow U^I{}_J \hY^J \label{hY-transform}
\ee
under the $T$-duality transformations (\ref{T-duality}), and fulfill
\be
\hY^I \eta_{IJ} \hY^J = 0 \label{hY-square}
\ee
identically on ${\cal M}_V$. Furthermore, we will need an additional
quantity, the invariant dilaton $S^{\rm inv}$, that transforms as
\be
S^{\rm inv} \rightarrow S^{\rm inv} + b
\ee
under $S$-duality and is invariant under $T$-duality. The motivation
is that these properties characterize the $\hX^I$ and the
perturbatively
invariant dilaton in the semi-classical case, i.e. when $f = f_{0}$.
The new
coordinates $Y^I$ are then defined as
\bea
Y^0 & = & \hY^0 \cr
Y^1 & = & \hY^0 S^{\rm inv} \cr
Y^i & = & \hY^i .
\eea
We can of course also work with the inhomogeneous covariant
coordinates $S^{\rm inv}$ and $(T^{\rm cov})^i = Y^i / Y^0$, which
parameterize ${\cal M}_V$.

We begin by regarding the $\hX^I$ as homogeneous coordinates on
${\cal M}_V$. We introduce the notation $\hpa_I = {\partial \over
\partial \hX^I}$. The jacobian matrix and its inverse for the
transformation
$X^{I}\rar\hX^{I}$
are
\bea
\frac{\pa\hX^{I}}{\pa X^{J}} = \de^{I}_{J} + \de^{I}_{1}B_{J}\,, &&
\frac{\pa
X^{I}}{\pa \hX^{J}} = \de^{I}_{J} - \de^{I}_{1}\frac{B_{J}}{1 -
B_{1}}\,.
\eea
Here, $B_{J} = 2\eta_{JK}\frac{\hX^{K}}{\hX^{0}} + \pa_{J}\pa_{1}f +
\de^{0}_{J}\frac{\pa_{0}f}{X^{0}}$. Hence, the change of coordinates
{}from $X^I$
to $\hX^I$ is
singular whenever the jacobian determinant
\be
\det \left( \pa_I \hX^J \right) = \pa^{2}_{1}f
\ee
vanishes but is regular generically. It follows from the expansion
(\ref{Fourier}) of $f$ that an irreducible component of the singular
locus is given by the divisor $q = 0$, i.e. $S = i \infty$. Indeed,
the $\hX^I$ fulfill the relation $\hX^I \eta_{IJ} \hX^J = 0$ at $q =
0$ and are thus not independent there. In a sense, there is a
physical singularity in the `weak coupling limit' $q = 0$, and
encircling this divisor produces exactly the $S$-duality monodromy
transformation (\ref{theta_shift}). Although a prepotential does not
exist in
the semi-classical limit \cite{Ceresole95} (essentially because the
transformation $X\rar\hX$ is singular in this limit) a prepotential
generically exists. In fact, $\hF = \frac{1}{2}\hX^{I}\hF_{I}$ has the
property $\hF_{I} = \hpa_{I}\hF$.

We now define
\be
\tilde{S} = {1 \over 2 (n + 2)} \eta^{IJ} \hpa_I \hF_J \label{tildeS}
\ee
and
\be
\hY^I = {1 \over 2} \left( \hX^I - \sqrt{-{\hX^K \eta_{KL} \hX^L
\over  \hpa_M \tilde{S} \eta^{MN} \hpa_N \tilde{S}}} \eta^{IJ} \hpa_J
\tilde{S} \right) . \label{hY-def}
\ee
It follows from (\ref{theta_shift}) and (\ref{T-duality}) that
$\tilde{S}$ transforms under $S$- and $T$-duality as
\be
\tilde{S} \rightarrow \tilde{S} + b
\ee
and
\be
\tilde{S} \rightarrow \tilde{S} + {1 \over 2 (n + 2)} \eta^{IJ}
(\Lambda_U)_{IJ} , \label{barS-shift}
\ee
whereas the $\hY^I$ are invariant under $S$-duality, transform as
(\ref{hY-transform}) under $T$-duality and satisfy the constraint
(\ref{hY-square}). (Here we have used that $\hX^I \hpa_I \tilde{S} =
X^I \pa_I \tilde{S} = 0$). Expressing $\tilde{S}$ in terms of $S$ and
$f$ we get ($\triangle = \eta^{ij}\pa_{i}\pa_{j}$)
\be
\tilde{S} = S + {1 \over 2 (n + 2)} \left( \triangle f - {\pa_i \pa_1 f
\eta^{ij} \pa_j\pa_1 f \over \pa^{2}_1
f} - {4 \pa_1 f \over X^0 \pa^{2}_{1}f } \right) .
\ee
Note that $\tilde{S}$ will have singularities on the rational quadratic
divisors. A direct calculation gives
\be
\tilde{S} = S + {1 \over 2 (n + 2)} \left( \triangle f_0 - {4 \over 2
\pi i}
\right) + {\cal O}(q) ,
\label{Stilde}
\ee
so $\tilde{S}$ is regular in the semi-classical limit and generalizes
the
perturbatively pseudo-invariant dilaton
that has been discussed in the literature \cite{deWit95}. Furthermore,
\be
\hpa_I \tilde{S} = - {2 \over (2 \pi i)^2 q f_1} \left( \eta_{IJ} \hX^J
+ {\cal O}(q) \right)
\ee
and
\be
-{\hX^K \eta_{KL} \hX^L \over  \hpa_M \tilde{S} \eta^{MN} \hpa_N
\tilde{S}} = \left( {(2 \pi i)^2 q f_1 \over 2} \right)^2 + {\cal
O}(q^3) ,
\ee
so if we define the branch of the square root function in
(\ref{hY-def}) by
\be
\sqrt{-{\hX^K \eta_{KL} \hX^L \over  \hpa_M \tilde{S} \eta^{MN} \hpa_N
\tilde{S}}} = {(2 \pi i)^2 q f_1 \over 2} + {\cal O}(q^2)
\ee
we get
\be
\hY^I = \hX^I + {\cal O}(q) .
\ee
The invariant dilaton can now be written as
\be
S^{\rm inv} = \tilde{S} + {1 \over 2 (n + 2)} L , \label{Sinv}
\ee
where the function $L$ is of degree zero and is determined up to an
additive constant by the requirements that $S^{\rm inv}$ should have no
singularities
on ${\cal M}_V$, transform correctly under $S$-duality and be
invariant under $T$-duality. $L$ should thus be invariant under
$S$-duality and transform as
\be
L \rightarrow L - \eta^{IJ} (\Lambda_U)_{IJ}
\ee
under $T$-duality. We see that
\bea
\Psi = \exp \left(2 \pi i L \right)
\eea
is invariant under $T$-duality because of the integrality of the
$\Lambda_U$ matrices. Introducing $q^{\rm inv} = e^{2 \pi i S^{\rm
inv}}$, we  can expand $\Psi$ as
\bea
\Psi &=& \sum_{k = 0}^\infty \left(q^{\rm inv} \right)^k \Psi_k ,
\label{Psi}
\eea
where the functions $\Psi_k$ are independent of $Y^1$ and of
degree zero in $Y^0$ and the $Y^i$.

\setcounter{equation}{0}
\section{Automorphic properties of the prepotential}
In this section, we will use the invariant dilaton and the
covariant coordinates $\hY^I$ to
express the functions $f_k$, that enter in the expansion
(\ref{Fourier}) of the prepotential, in terms of another set of
functions with simple transformation properties under $T$-duality.

We will begin by reviewing the definition of an automorphic form of
$SO(2, n; {\bf Z})$, see e.g. \cite{Borcherds}. Define the domain
$P \subset \Gamma^{2, n} \otimes {\bf C}$ through
\be
P = \{ \hY^I \; | \; \hY^I \eta_{IJ} \hY^J = 0, \;
\hY^I \eta_{IJ} \overline{\hY^J} < 0 \} .
\ee
The $T$-duality group $SO(\Gamma^{2, n}) \cong SO(2, n; {\bf Z})$
has a natural action on $P$. An automorphic form of weight $w$ is now
defined as a homogeneous function on $P$ of degree $-w$ that is
invariant under $SO(2, n; {\bf Z})$. Alternatively, we can construct
a model of the Hermitian symmetric space of $SO(2,n)$
\be
H \cong {SO(2, n) \over SO(2) \times SO(n)}
\ee
as $H \cong P / C^*$, where $\lambda \in C^*$ acts as $\hY^I \mapsto
\lambda \hY^I$. $P$ is then a principal $C^*$-bundle over $H$ and an
automorphic form of weight $w$ is an $SO(2, n; {\bf Z})$ invariant
holomorphic section of $P^w$.

We have in fact already introduced a set of automorphic
forms of weight zero, or automorphic functions, namely the $\Psi_k$ of
(\ref{Psi}) if we regard them as
functions defined on $P$ rather than as functions of the $Y^I$ by using
the relations
\bea
\hY^0 & = & Y^0 \cr
\hY^1 & = & {Y^i \eta_{ij} Y^j \over Y^0} \cr
\hY^i & = & Y^i . \label{P-coordinates}
\eea

We can now easily construct more examples of automorphic forms by
expressing an arbitrary $S$- and $T$-duality invariant quantity
${\cal A}$ of degree $-w$ in terms of the covariant coordinates $Y^I$.
Invariance of ${\cal A}$ under $S$-duality means that it can be
expanded as
\be
 {\cal A} = \sum_{k = 0}^\infty (q^{\rm inv})^k A_k
\ee
where the functions $A_k$ are independent of $Y^1$ and of degree $-w$
in $Y^0$ and the $Y^i$. (We have assumed that ${\cal A}$ is regular in
the $S \rightarrow i \infty$ limit.) Invariance of ${\cal A}$ under
$T$-duality
is now tantamount to the $A_k$ being automorphic forms of weight $w$
when regarded as functions on $P$ via (\ref{P-coordinates}).

A simple example of such an invariant is $\Upsilon= \hX^I \eta_{IJ}
\hX^J = -X^{0}\pa_{1}f$, which is of degree $2$. It can be expanded in
terms of
the functions
$f_k$ in (\ref{Fourier}) as
\be
\Upsilon = - 2 \pi i \sum_{k = 1}^\infty k q^k f_k . \label{H-f}
\ee
and in terms of the invariant dilaton and functions $h_k$ of the
covariant coordinates as
\be
 \Upsilon = -2\pi i\sum_{k = 1}^\infty k(q^{\rm inv})^k h_k ,
\label{H-h}
\ee
Comparing the two expansions (\ref{H-h}) and (\ref{H-f}) of $\Upsilon$
and using the expression (\ref{Sinv}) with (\ref{tildeS})
for the invariant dilaton and (\ref{hY-def}) for the covariant
coordinates, we can express the $f_k$ in terms of the $h_k$ and vice
versa.

In this way we get
\bea
h_1 & = & \exp \left( - {2 \pi i \over 2 (n + 2)} [ \triangle f_0
- {4 \over 2 \pi i} + L_0 ] \right) f_1 \cr
h_2 & = & \frac{n}{n+2}\exp\left( - {4 \pi i \over 2 (n + 2)} [
\triangle f_0 - {4 \over 2 \pi i} + L_0 ] \right) \Bigl[ f_{2} -
\frac{2\pi i}{4n}f_{1}\triangle f_{1}  - \frac{2\pi i}{4n}f_{1}L_{1}
  \non \\ && +
\frac{2\pi i}{16n}(2-n)\pa_{i}f_{1}\eta^{ij}\pa_{j}f_{j} + \frac{(2\pi
i)^{2}}{16n}f_{1}(\pa_{i}f_{1} \non - \frac{2\pi
i}{2(n+2)}f_{1}\pa_{i}\triangle f_{0})\eta^{ij}\pa_{j}L_{0}
 \non \\  &&  +
\frac{(2\pi i)^{2}}{16n}f_{1}\pa_{i}f_{1}\eta^{ij}\pa_{j}\triangle
f_{0} -
\frac{(2\pi i)^{3}}{64n(n+2)}f_{1}^{2}\pa_{i}\triangle f_{0}
\eta^{ij}\pa_{j}\triangle f_{0} \Bigr]   \cr
& \ldots
\eea
where the $L_k$'s are defined through
\be
L = \sum_{k=0}^{\infty} L_k q^k \,.
\ee
The expressions for the $h_k$ become increasingly complex for larger
$k$, but the general structure is that $h_k$ equals $\exp \left( - {2
\pi i k \over 2 (n + 2)} \left( \eta^{ij} \pa_i \pa_j f_0 - {4 \over
2 \pi i} + L_0 \right) \right)$ times a differential polynomial in
the $f_{k^\prime}$ and $L_{k^\prime}$ of degree two and charge $k$ if
$f_{k^\prime}$ and $L_{k^\prime}$ are assigned charge $k^\prime$.
A heuristic argument goes as follows: First we
observe that at any given order in the $q$ expansion of (\ref{H-h}),
only $f_{1}$
appears in the denominator. Hence generically $h_{k}$ should on the face
of it
be singular when $f_{1} = 0$. However, $\Upsilon = -X^{0}\pa_{1}f$, and
is regular at the zeros of $f_{1}$ so generically (assuming there is no
great conspiracy) this should be true also term by term in the
expansion.
Furthermore, $f$
can be expanded (by comparing with the type II expression) in
trilogarithms,
and the first derivative of a trilogarithm is non-singular (except at
infinity); in particular nothing special should happen when $f_{1}=0$.
Hence, we conclude that terms which are singular when $f_1 = 0$ 
are absent.

Since $\hpa_{I_{1}}\cdots\hpa_{I_{k}}\hF$ transforms as a tensor it can
be
used together with $\hX^{I}$ to build further invariants.
These can also be expanded in terms of automorphic forms
in the same way as above.
These are not independent, though, but can be expressed in
terms of the $h_k$ above and certain automorphic combinations of
their derivatives. The simplest such combination is the Schwarzian
derivative
\be
h \triangle h - {n + 2 \over 4} \eta^{ij} \pa_i h \pa_j h
,
\ee
which is a meromorphic automorphic form of weight $-2$ whenever $h$ is.

The above construction goes through unaltered even if we set $L_k =
0$ for $k > 0$. The only difference is that $S^{\rm inv}$ will have
singularities somewhere on ${\cal M}_V$, but we can still construct
meromorphic automorphic forms $h_k$ in terms of $\Psi_0$ and the $f_k$.
We can go further and also set $L_0 = 0$. The $h_k$ will then transform
with the phase $\exp \left( - {2 \pi i k \over 2 (n + 2)} \eta^{IJ}
(\Lambda_U)_{IJ} \right)$, but products of the form $h_{k_1} \cdot
\ldots \cdot h_{k_m}$ such that $k_1 + \ldots + k_m = 2 (n + 2)$ will
be true meromorphic automorphic forms due to the integrality of the
$\Lambda_U$ matrices. Following the path outlined above we  thus get
explicit formulas relating the functions $f_k$ to automorphic forms.

\bigskip\bigskip\noindent{\Large\bf Acknowledgements}

\bigskip\noindent

The authors acknowledge useful discussions with G.~Cardoso, J.~Louis,
P.~Mayr
and S.~Theisen. PB also thanks the Theory Division at CERN and the
Mittag-Leffler Institute, Stockholm where some of this work was carried
out. NW would like to thank the CERN Theory Division for
hospitality. The work of PB was  supported in part by the National
Science Foundation grant NSF  PHY94-07194.

\begin{appendix}
\setcounter{equation}{0}
\section{The low-dimensional cases}
Our discussion so far applies to $SO(2, n; {\bf Z})$ for arbitrary
$n$. In this appendix, we will make some comments on the cases when
$n = 1, 2, 3$. Here we have a more explicit knowledge of the space of
automorphic forms, which allows for quite explicit calculations to
be performed.

\subsection{$n = 1$}
Here we use the isomorphism $SO(2, 1; {\bf Z}) \cong PSL(2; {\bf
Z})$ given by
\be
PSL(2; {\bf Z}) \ni \pmatrix{a & b \cr c & d} \mapsto U = \pmatrix{
d^2 & c^2 & 2 c d \cr b^2 & a^2 & 2 a b \cr b d & a c & a d + b c}
\in SO(2, 1; {\bf Z}) .
\ee
The action of the $T$-duality transformation $\hX^I \rightarrow
U^I{}_J \hX^J$ on the inhomogeneous special coordinate $T = X^2 / X^0
= \hX^2 / \hX^0$ is then
\be
T \rightarrow {a T + b \over c T + d} ,
\ee
i.e. the standard $PSL(2, {\bf Z})$ action on the upper halfplane.
Classically, the vector multiplet moduli space ${\cal M}_V$ is thus
given by the dilaton factor times the $PSL(2, {\bf Z})$ fundamental
domain. The only rational quadratic divisor is $T = i$, at which
point a $U(1)$ factor gets enhanced to $SU(2)$ without any additional
hyper multiplet doublets becoming massless.

Writing an automorphic form $\Theta$ of weight $w$ in the form
\be
\Theta = (X^0)^{-w} \theta (T) ,
\ee
we see that automorphicity of $\Theta$ is equivalent to $\theta$
being an $PSL(2, {\bf Z})$ modular form of weight $2 w$, i.e.
\be
\theta \left( {a T + b \over c T + d} \right) = (c T + d)^{2 w}
\theta_k (T) .
\ee
(Note the different conventions for the weights of $SO(2, 1; {\bf
Z})$ automorphic forms and $PSL(2, {\bf Z})$ modular forms.)

The graded ring of holomorphic automorphic forms is freely generated
by the Eisenstein series $E_4 (T)$ and $E_6 (T)$ of weights $4$ and
$6$ respectively. They have simple zeros at $T = \rho = \exp i \pi /
3$ and $T = i$ respectively. We will also have use for the weight $0$
form $j (T) = 1728 E_4^3 (T) / (E_4^3 (T) - E_6^2 (T))$, which is a
one-to-one mapping from the fundamental domain to the Riemann sphere
such that $j (T)$ has a simple pole at the cusp $T = i \infty$ and a
triple zero at $T = \rho$ and $j (T) - 1728$ has a double zero at $T
= i$. The function $L_0$ should have logarithmic singularities at the
rational quadratic divisor $T = i$ and also at the cusp $T = i \infty$;
it is thus proportional to $\log (j(T) - 1728)$.

A model of this kind can be obtained by compactifying the $E_8 \times
E_8$ heterotic string on $K3 \times T^{2}$, restricting to the subspace
where the complexified K\"ahler modulus $U$ of the $T^{2}$ equals its
complex structure modulus $T$ so that an extra $SU(2)$ group arises
and breaking all non-abelian gauge symmetry by embedding $10$
instantons in each $E_8$ factor and $4$ instantons in the $SU(2)$
factor. This model is dual to the type IIA string compactified on a
degree $12$ hypersurface in the weighted projective space
$\IP^4_{1, 1, 2, 2, 6}$
with Hodge numbers $h^{1, 1} = 2$ and $h^{2,
1} = 128$ \footnote{This pair of dual models has been studied
extensively
\cite{Kachru-Vafa,Klemm95,KLT}.}. A particular feature of this
Calabi-Yau
space is the existence
of a certain symmetry
between the $S$ and $T$ moduli \cite{Candelas,hkty} which, together
with
the singularity
structure described above, uniquely determines the special geometry
\cite{Henningson2}.

\subsection{$n = 2$}
Here we use that $SO(2, 2; {\bf Z})$ is generated by two commuting
$PSL(2; {\bf Z})$ subgroups with the standard action on the
inhomogeneous special coordinates
\be
T={X^2-X^3\over X^0}\,,\;\; U={X^2+X^3\over X^0},
\ee
and a ${\bf Z}_2$ transformation that exchanges them. Classically,
${\cal M}_V$ is given by the dilaton factor,
parametrized as usual by $S = X^1/ X^0$,
 times the product of two
$PSL(2; {\bf Z})$ fundamental domains modulo the operation that
exchanges them. The only rational quadratic divisor is $T - U = 0$,
where an $SU(2)$ factor arises without any extra hyper multiplets.
(The points $T = U = i$ and $T = U = \rho$ are special in that the
divisor intersects its images under $SO(2, 2; {\bf Z})$; at these
points we instead get an $SU(2) \times SU(2)$ or $SU(3)$ factor
respectively.)

Writing an automorphic form $\Theta$ of weight $w$ as
\be
\Theta = (X^0)^2 \theta (T, U) ,
\ee
we get that $\theta$ is symmetric in $T$ and $U$ and has weight $w$
with respect to both $PSL(2; {\bf Z})$ factors.

The graded ring of holomorphic automorphic forms is generated by $E_4
(T) E_4 (U)$, $E_6 (T) E_6 (U)$ and $E_4 (T) E_6 (U) + E_4 (U) E_6
(T)$. We will also have use for $j (T) - j (U)$, which has a
simple zero on the rational quadratic divisor $T - U = 0$ and a simple
pole at $T = i \infty$ and $U = i \infty$. In fact, $L_0$ is
proportional to $\log (j (T) - j (U))$.

There exists three models  of this type (two of which are isomorphic)
which can
be obtained by  compactifying  the $E_8 \times
E_8$ heterotic string on $K3 \times T^{2}$ and embedding $12-l$ and
$12+l$
instantons in the first and second $E_8$ factor respectively with 
$l=0,1,2$ \cite{Kachru-Vafa,AFIQ,DMW}.
It is then possible to break all the non-abelian gauge symmetries,
leaving only a
$U(1)^4$ gauge symmetry from the $N_V=3$ vector multiplets and the
graviphoton (inherited from the $T^2$ and the dilaton).
These models are dual to type IIA compactifications on a particular
kind of
Calabi-Yau threefolds, elliptic fibrations over a so called Hirzebruch
surface
$F_l$, where the latter is itself a $\IP^1$ fibration over a base
$\IP^1$ \cite{Morrison}.
 It has been shown that the cases $l=0$ and $l=2$ are
isomorphic \cite{Morrison,phases} and we will thus only consider $l=0,1$.

We will now study certain properties of the prepotential $F$, in
particular the
contribution from the one-instanton correction, $f_1$. Recall, that $F$
can
be written as
\be
F = {X^1 \over X^0} X^i \eta_{ij} X^j + \sum_{k = 0}^\infty q^k f_k .
\label{prepotentialapp}
\ee
Along the lines of the discussion in section~6 one can show that $f_k$
can be
written in the form
\be
f_k = \exp \left( 2 \pi i k {1 \over 2 (n + 2)} \triangle h_0 \right)
\left( h_k + \ldots \right) .
\ee
The functions $h_k(X)$ are independent of $X^1$ and of degree two in
$X$ and transform (for $k > 0$) as automorphic forms up to a phase in
the sense that
$$
h_k(\tilde{X}) = \exp \left( -2 \pi i {k \over 2 (n + 2)} \eta^{IJ}
\Lambda_{IJ} \right) h_k(X) .
$$
The ellipsis in the formula for $f_k$ denote a polynomial in
$\triangle^{m} h_{k^\prime}$ except $h_0$ and $\triangle h_0$.

Let us now consider the cases of interest. We introduce the
inhomgeneous coordinates
$S$, $T$ and $U$,
which are the natural coordinates from the heterotic compactification,
as well as
\be
q_S = q = {\rm exp}(2\pi i S)\,,\;\; q_T  = {\rm exp}(2\pi i T)\,,
\;\; q_U  = {\rm exp}(2\pi i U).
\ee
We can then rewrite (\ref{prepotentialapp}) as
\be
F = STU + F_{\rm 1-loop}(T,U) + \sum_{k=1}^{\infty} q^k f_k(T,U)
\ee
where
\be
F_{\rm 1-loop} = F_{\rm cubic} +
{1\over (2\pi i)^3}\sum_{i,j}n(i,j,0) q_T^j q_U^{i-j}
\ee
and
\be
F_{\rm 1-inst}= f_1 = {1\over (2\pi
i)^3}\sum_{i,j=0}^{\infty}n(i,j,1)q_T^j
q_U^{i-j-1} q_S .
\ee
Note that in order for the above expansion to be valid we take
${\rm Im} U< {\rm Im} T <{\rm Im} S$.
We have here used the duality between the heterotic string on $K3\times
T^2$
and the dual type IIA Calabi-Yau manifold in order to express the
1-loop and
1-instanton corrected prepotential in terms of the number of rational
curves
$n(i,j,k)$ \footnote{The $n(i,j,k)$ have been computed for the models 
in question using INSTANTON \cite{hkty}.}. Although $n(i,j,k)$ does 
in general depend on the
particular manifold, it turns out that for the two cases that we are
studying,
$l=0,1$, when restricted to the 1-loop corrections, they agree, i.e.
$n(i,j,0)$ are the same for the two models.
Furthermore, $F_{\rm cubic}$ is
given by
\be
F_{\rm cubic}={1\over 3} U^3 - \delta_{l1}\left({1\over 2} U^2 T -
{1\over 2} U T^2\right)
\ee
where $\delta_{l1}$ is the usual Kronecker delta.
The cubic couplings in the two theories can however be made to agree
if the dilaton in the $l=1$ case is shifted by
\be
S\to S +{1\over 2} U -{1\over 2} T\, .
\ee
Clearly, this shift will remove the mixed terms in $U$ and $T$ while
leaving the rest of the 1-loop effect intact. However, the  instanton
corrections will pick up non-integer exponents of the type
$(q_U q_T)^{-1/2 k}$ for the $k$-th order space-time instanton
correction. Thus,
although it seems to be a perfectly valid transformation from a
perturbative
point of view, we come to the conclusion that the shift of the dilaton
is not
allowed non-perturbatively. Hence, the two models are inequivalent.

By using the expression of $h_1$ in terms of $f_1$ and $f_0$,
\be
h_1={\rm exp}(-\pi i\pa_U\pa_T f_0) f_1
\ee
it is indeed possible to write $h_1$ in terms of the generators of the
graded ring of holomorphic forms;
\be
h_1^{l=0}={-2\over 2 \pi i} {E_4(q_T)E_4(q_U) \over (j(T)-j(U))
\eta^{12}(q_T) \eta^{12}(q_U)} \Big({E_6(q_T)\over
\eta^{12}(q_T)}\Big)\Big({E_6(q_U)\over
\eta^{12}(q_U)}\Big)
\ee
and
\be
h_1^{l=1}={1\over 2 \pi i} {E_4(q_T)E_4(q_U) \over (j(T)-j(U))
\eta^{12}(q_T) \eta^{12}(q_U)}\Big\{ \Big({E_6(q_T)\over
\eta^{12}(q_T)}\Big)^2+\Big({E_6(q_U)\over
\eta^{12}(q_U)}\Big)^2\Big\}
\ee
Note that these functions indeed transform with a phase under
the $SO(2, 2; {\bf Z})$ transformation that exchanges $T$ and $U$.

\subsection{$n = 3$}
Here we use that $SO(2, 2; {\bf Z}) \cong Sp(4; {\bf Z})$.
Assembling the inhomogeneous special coordinates
\be
T={X^2-X^3\over X^0},\, U={X^2+X^3\over X^0}, \, V = {X^4 \over X^0}
\ee
into $\tau = \pmatrix{ T & V \cr V & U}$,
the action of $\pmatrix{A & B \cr C & D} \in Sp(4; {\bf Z})$ is
\be
\tau \rightarrow {A \tau + B \over C \tau + D} ,
\ee
which is the standard $Sp(4; {\bf Z})$ on the Siegel generalized
upper halfplane. There are two different rational quadratic divisors
\cite{NG}:
In addition to $T - U = 0$ (which only gives an $SU(2)$ factor) there
is also $V = 0$ where we also get some model-dependent number of
extra hyper multiplets \cite{Cardoso1,Cardoso2}.
In the $Sp(4)$ context, these divisors are
known as the Humbert surfaces $H_4$ and $H_1$ respectively.
Automorphic forms then correspond to Siegel modular forms for $Sp(4;
{\bf Z})$. It follows from the Koechner boundedness principle that a
nearly  holomorphic automorphic form is in fact holomorphic. The ring
of such forms is generated by the Eisenstein series $E_4 (T,
U, V)$, $E_6 (T, U, V)$, $E_{10} (T, U, V)$, and $E_{12} (T, U, V)$
together with the cusp form ${\cal C}_{35} (T, U,
V)$ \cite{Igusa}.

\end{appendix}

\def\href#1#2{#2}

\begingroup\raggedright\endgroup

\end{document}